\pdfoutput=1 
\documentclass[structabstract]{aa}  
\usepackage{graphicx}
\usepackage{txfonts}
\usepackage{natbib}
\begin{document}
\title{Fragmentation and dynamical collapse of the starless high-mass star-forming region IRDC\,18310-4\thanks{Based on observations carried out with the IRAM
    Plateau de Bure Interferometer. IRAM is supported by INSU/CNRS
    (France), MPG (Germany) and IGN (Spain). The data are available in
    electronic form at the CDS via anonymous ftp to
    cdsarc.u-strasbg.fr (130.79.128.5) or via
    http://cdsweb.u-st4rasbg.fr/cgi-bin/qcat?J/A+A/}.}


   \author{H.~Beuther
          \inst{1}
          \and
          H.~Linz
          \inst{1}
          \and
          J.~Tackenberg
          \and
          Th.~Henning
           \inst{1}
          \and
          O.~Krause
          \inst{1}
          \and
          S.~Ragan
          \inst{1}
          \and
          M.~Nielbock
          \inst{1}
          \and
          R.~Launhardt
          \inst{1}
          \and
          S.~Bihr
          \inst{1}
          \and
          A.~Schmiedeke
          \inst{1,2}
          \and
          R.~Smith
          \inst{3}
          \and
          T.~Sakai
          \inst{4}
           }
   \institute{$^1$ Max-Planck-Institute for Astronomy, K\"onigstuhl 17,
              69117 Heidelberg, Germany, \email{name@mpia.de}\\
              $^2$ University of Cologne, Z\"ulpicher Strasse 77, 50937 K\"oln, Germany\\
              $^3$ Institute for Theoretical Astrophysics, Univ.~of Heidelberg,
Albert-Ueberle-Str. 2, 69120 Heidelberg, Germany\\
$^4$ Institute of Astronomy, The University of Tokyo, Osawa, Mitaka, Tokyo 181-0015, Japan
}



\abstract
{Because of their short evolutionary time-scales, the earliest stages
  of high-mass star formation prior to the existence of any embedded
  heating source have barely been characterized until today.}
{We study the fragmentation and dynamical properties of a massive
  starless gas clump at the onset of high-mass star formation.}
{Based on Herschel continuum data we identify a massive gas clump 
  that remains far-infrared dark up to 100\,$\mu$m wavelengths. The
  fragmentation and dynamical properties are investigated by means of
  Plateau de Bure Interferometer and Nobeyama 45\,m single-dish
  spectral line and continuum observations.}
{The massive gas reservoir (between $\sim$800 and
  $\sim$1600\,M$_{\odot}$, depending on the assumed dust properties)
  fragments at spatial scales of $\sim$18000\,AU in four cores.
  Comparing the spatial extent of this high-mass region with
  intermediate- to low-mass starless cores from the literature, we
  find that linear sizes do not vary significantly over the whole mass
  regime. However, the high-mass regions squeeze much more gas into
  these similar volumes and hence have orders of magnitude larger
  densities. The fragmentation properties of the presented low-to
  high-mass regions are consistent with gravitational instable Jeans
  fragmentation. Furthermore, we find multiple velocity components
  associated with the resolved cores.  Recent radiative transfer
  hydrodynamic simulations of the dynamic collapse of massive gas
  clumps also result in multiple velocity components along the line of
  sight because of the clumpy structure of the regions.  This result
  is supported by a ratio between viral and total gas mass for the
  whole region $<$1.}
{This apparently still starless high-mass gas clump exhibits clear
  signatures of early fragmentation and dynamic collapse prior to the
  formation of an embedded heating source. A comparison with regions
  of lower mass reveals that the linear size of star-forming regions
  does not necessarily have to vary much for different masses,
  however, the mass reservoirs and gas densities are orders of
  magnitude enhanced for high-mass regions compared to their
  lower-mass siblings.}  
   \maketitle

\section{Introduction}
\label{intro}

Independent of the various formation scenarios for high-mass stars
that are discussed extensively in the literature (e.g.,
\citealt{zinnecker2007,beuther2006b}), the initial conditions required
to allow high-mass star formation at all are still poorly
characterized. The initial debate even ranged around the question
whether high-mass starless gas clumps should exist at all, or whether
the collapse of massive gas clumps starts immediately without any
clear starless phase in the high-mass regime (e.g.,
\citealt{motte2007}). Recent studies indicate that the time where
massive gas clumps exist without embedded star formation is relatively
short (on the order of 50000\,yrs), but nevertheless, high-mass
starless gas clumps do exist (e.g.,
\citealt{russeil2010,tackenberg2012}).

Current questions in that field are: Are high-mass gas clumps
dominated by a single fragment or do we witness already strong
fragmentation during earliest evolutionary stages (e.g.,
\citealt{bontemps2010})? What are the kinematic properties of the gas?
Are the clumps sub- or super-virial? Do we see streaming motions
indicative of turbulent flows (e.g.,
\citealt{bergin2004,vazquez2006,heitsch2008,banerjee2009})?

To address these questions high spatial resolution is mandatory.
Furthermore, to study the dense gas, dust continuum as well as
spectral line observations are required. Therefore we tackle these
questions via a dedicated study of the genuine high-mass starless gas
clump IRDC\,18310-4 that was first identified by
\citet{sridharan2005}. Their early study based on 1.2\,mm continuum
emission, 8\,$\mu$m extinction and NH$_3$ emission revealed an
infrared dark gas clump with a mass of $\sim 840$\,M$_{\odot}$, a
temperature of $\sim 18$\,K, and a NH$_3$(1,1) line-width measured
with the Effelsberg 100\,m telescope with 40$''$ of $\sim
1.7$\,km\,s$^{-1}$. The kinematic distance of the source is $\sim
4.9\pm 0.26$\,kpc (\citealt{ragan2012b}, recalculated following
\citealt{reid2009}) and its velocity of rest $v_{\rm{lsr}}\sim
86.5$\,km\,s$^{-1}$ \citep{sridharan2005}.

\begin{figure*}[htb] 
\caption{Compilation of the continuum data from 70\,$\mu$m to
  500\,$\mu$m wavelength in color-scale as labeled in each panel. The
  scale is chosen in each image individually to highlight the most
  important features. Contour levels of the 870\,$\mu$m data start at
  the 3$\sigma$ levels of 0.189\,mJy\,beam$^{-1}$ and continue in in
  3$\sigma$ steps. The white small box in the top-left panel marks the
  smaller region discussed in the remaining figures. A scale-bar is
  shown in the bottom-left panel.}
\label{overview}
\end{figure*}

The target region is at a projected separation of $\sim 3.4'$ (at the
given distance corresponding to $\sim 5.1$\,pc) from the luminous
high-mass protostellar object (HMPO) IRAS\,18310-0825
\citep{sridha,beuther2002a} but otherwise in no outstanding part of
our Galaxy. This region is part of the Herschel EPoS key project
(Earliest Phases of Star Formation) that studies a large sample of high-
and low-mass star-forming regions at the earliest evolutionary stages
(\citealt{ragan2012b}, \citealt{launhardt2013},
\citealt{henning2010}, \citealt{linz2010},
\citealt{beuther2010b,beuther2012a}, \citealt{stutz2010}).

Except for these characterizations, little additional information exist
for that region. Here, we present Herschel far-infrared continuum
observations as well as single-dish and interferometrically obtained
N$_2$H$^+$(1--0) observations from the Nobeyama 45\,m telescope and
the Plateau de Bure Interferometer. The latter data also allow us to
derive high-spatial-resolution 3\,mm continuum data.

\section{Observations} 
\label{obs}

\subsection{Herschel}

The cloud complex with a size of $\sim6'\times 6'$ was observed with
PACS \citep{A&ASpecialIssue-PACS} on Herschel
\citep{A&ASpecialIssue-HERSCHEL} on April 19th, 2011 (PACS OBS ID
134221906[0-3]).  Scan maps in two orthogonal directions with scan leg
lengths of $\sim 10'$ were obtained with the medium scan speed of
$20''$/s.  The raw data have been reduced up to level-1 with the HIPE
software \citep{A&ASpecialIssue-PACS,ott2010}, version 6.0, build
1932. Besides the standard steps, we applied a 2$^{nd}$level
deglitching, in order to remove bad data values from a given pixel map
by $\sigma$-clipping the flux values which contribute to each pixel.
We used the time-ordered option and applied a 25$\sigma$ threshold.
The final level-2 maps were processed using Scanamorphos version 8
\citep{roussel2012}. Since the field of view contains bright emission
on scales larger than the map, we applied the ``galactic'' option and
included the non-zero-acceleration telescope turn-around data.  The
flux correction factors provided by the PACS ICC team were applied.
The 70\,$\mu$m extinction feature and the PdBI continuum and line data
exhibited a small offset of approximately $3''$ in R.A. and $1''$ in
Declination. Although the interferometer positional accuracy is
usually superior because of the extremely well-known positions of the
reference quasars, we double-checked the 70\,$\mu$m PACS data via a
comparison to the Spitzer MIPSGAL 24\,$\mu$m survey which has also a
very high spatial accuracy \citep{carey2009}. Again, the same offset
between MIPSGAL 24\,$\mu$m and Herschel 70\,$\mu$m was found.  These
Herschel data were taken at still an early Herschel time, and
positional offsets on this order are known during that time.
Therefore, we shifted the 70\,$\mu$m map by this little offset in the
high-resolution overlays from Figure \ref{zoom} onwards.
The spatial resolution of the 70, 100 and 160\,$\mu$m data is $5.6'',
6.8''$ and $11.4''$, respectively.

Maps at 250, 350, and 500\,$\mu$m were obtained with SPIRE
\citep{A&ASpecialIssue-SPIRE} on March 11th, 2010 (SPIRE OBS ID
1342192067). Two times two $20'$ scan legs were used to cover the
source. The data were processed up to level-1 within HIPE, developer
build 5.0, branch 1892, calibration tree 5.1 using the standard
photometer script (POF5\_pipeline.py, dated 2.3.2010) provided by the
SPIRE ICC team.  The resulting level-1 maps have been further reduced
using Scanamorphos, version 9 (patched, dated 08.03.2011). This
version included again the essential de-striping for maps with less
than 3 scan legs per scan. In addition, we used the ``galactic''
option and included the non-zero-acceleration telescope turn-around
data. Corrections for the transition from point sources to extended
sources were applied according to the SPIRE Manual.  The spatial
resolution of the 250, 350 and 500\,$\mu$m data is $18.1'', 24.9''$
and $36.6''$, respectively.

\begin{table*}[htb]
\caption{Core parameters from the PdBI 3.2\,mm data}
\begin{tabular}{lrr|rr|rr|rr||rr|rr|rr}
\hline \hline
\# & R.A. & Dec. & $S_{\rm{peak}}$ & $S$ & N$_{\rm{H}_2}$ & $M$ & N$_{\rm{H}_2}$ & $M$ & $\Delta v_1$ & $\Delta v_2$ & $M_{\rm{vir}}$ & $M_{\rm{vir}}$ & $M_{\rm{vir}}$ & $M_{\rm{vir}}$ \\
   & (J2000.0) & (J2000.0) & $\left(\frac{\rm{mJy}}{\rm{beam}}\right)$ & (mJy) & $\left(\frac{10^{23}}{\rm{cm}^2}\right)$ & (M$_{\odot}$) & $\left(\frac{10^{23}}{\rm{cm}^2}\right)$ & (M$_{\odot}$) &  \multicolumn{2}{|c|}{(km\,s$^{-1}$)} &  \multicolumn{2}{|c|}{$\rho \propto \frac{1}{r}$, (M$_{\odot}$)} &  \multicolumn{2}{|c}{$ \rho\propto \frac{1}{r^2}$, (M$_{\odot}$)} \\
    &              &              &      &     & \multicolumn{2}{|c|}{MRN1977} & \multicolumn{2}{|c||}{OH94} & & & $v_1$ & $v_2$ & $v_1$ & $v_2$   \\ 
\hline
mm1 & 18:33:39.472 & -08:21:09.96 & 0.90 & 1.2 & 6.1 & 107  & 2.0 &  36 & 2.4 & 0.6 & 48 & 3 & 32 & 2 \\
mm2 & 18:33:39.277 & -08:21:09.83 & 0.75 & 0.9 & 5.1 &  81  & 1.7 &  27 & 2.4 & 0.8 & 48 & 5 & 32 & 4 \\
mm3 & 18:33:39.334 & -08:21:16.74 & 0.91 & 1.2 & 6.2 & 107  & 2.1 &  36 & 2.0 & 0.6 & 33 & 3 & 22 & 2 \\
mm4 & 18:33:40.013 & -08:21:04.84 & 0.47 & 0.6 & 3.2 &  54  & 1.1 &  18 & --  & --  &  & & & \\
\hline \hline
\end{tabular}
\footnotesize{Notes: MRN1977 \citet{mathis1977}, OH94
  \citealt{ossenkopf1994}, $v1$ and $v2$ correspond to the fitted peak
  velocities of the two components in Fig.~\ref{pdbi_nobeyama}.}
\label{fluxes}
\end{table*}

\subsection{Plateau de Bure Interferometer}

We observed IRDC\,18310-4 with the Plateau de Bure Interferometer
during five nights in October and November 2009 at 93\,GHz in the C
and D configurations covering projected baselines between
approximately 13 and 175\,m.  The observations were conducted in a
track-sharing mode together with the IRDC\,18454-1 in the neighborhood
of W43 \citep{beuther2012a}. The 3\,mm receivers were tuned to
92.835\,GHz in the lower sideband covering the N$_2$H$^+$(1--0) as
well as the 3.23\,mm continuum emission. At the given frequency, the
primary beam of the PdBI is $\sim 54''$. For continuum measurements we
placed six 320\,MHz correlator units in the band, the spectral lines
were excluded in averaging the units to produce the final continuum
image.  Temporal fluctuations of amplitude and phase were calibrated
with frequent observations of the quasars 1827+062 and 1829-106.  The
amplitude scale was derived from measurements of MWC349 and 3c454.3.
We estimate the final flux accuracy to be correct to within $\sim
15\%$. The phase reference center is R.A.~(J2000.0) 18:33:39.468 and
Dec.~(J2000.0) -08:21:07.65, and the velocity of rest $v_{\rm{lsr}}$
is 86.5\,km\,$s^{-1}$ for the two velocity components, respectively.
The data were imaged with a robust weighting scheme close to uniform
weighting. The synthesized beam of the line and continuum data is
$4.3''\times 3.0''$ with a P.A. of 14 degrees.  The $3\sigma$
continuum rms is 0.24\,mJy\,beam$^{-1}$. The $3\sigma$ rms of the
N$_2$H$^+$(1--0) data measured from an emission-free channel with a
spectral resolution of 0.2\,km\,s$^{-1}$ is 21\,mJy\,beam$^{-1}$. Data
calibration and imaging was performed with the GILDAS software
package\footnote{http://www.iram.fr/IRAMFR/GILDAS}.

\subsection{Nobeyama 45\,m telescope}

The N$_2$H$^+$ data has been observed using the BEARS receiver at the
NRO 45\,m telescope in Nobeyama, Japan, in April 2010. The receiver
has been tuned to $93.17346$\,MHz, covering the full hyperfine
structure of the N$_2$H$^+(1-0)$ transition with average system
temperatures of $T_{\rm{sys}}\sim 200$\,K. At this frequency the
telescope beam is $18.2''$ and the observing mode provides a spectral
resolution of $0.2$\,km\,s$^{-1}$ at a bandwidth of 32\,MHz. The
software package nostar \citep{sawada2008} was used for the data
reduction, sampling the data to a pixel size of $7''$ with a
spheroidal convolution and smoothing the spectral resolution to
$0.5$\,km\,s$^{-1}$. The rms of the final map is $\sim 0.11$\,K per
$0.5$\,km\,s$^{-1}$ wide channel. Strong winds hampered the pointing
for part of the observations, contributing to spatial uncertainties.

\subsection{Merging of PdBI and Nobeyama data}
\label{mergedata}

To combine the N$_2$H$^+(1-0)$ from the Nobeyama 45\,m with the PdBI
data, the Nobeyama data were converted into units of Jy, and then the
combination was done within the GILDAS software package by the task
{\sc uvshort}. The synthesized beam of the combined dataset is
$5.1''\times 3.2''$ with a P.A. of 8 degrees. The $3\sigma$ rms of the
merged N$_2$H$^+$(1--0) data measured from an emission-free channel
with a spectral resolution of 0.2\,km\,s$^{-1}$ is
54\,mJy\,beam$^{-1}$.

\begin{figure}[htb] 
\includegraphics[width=10cm]{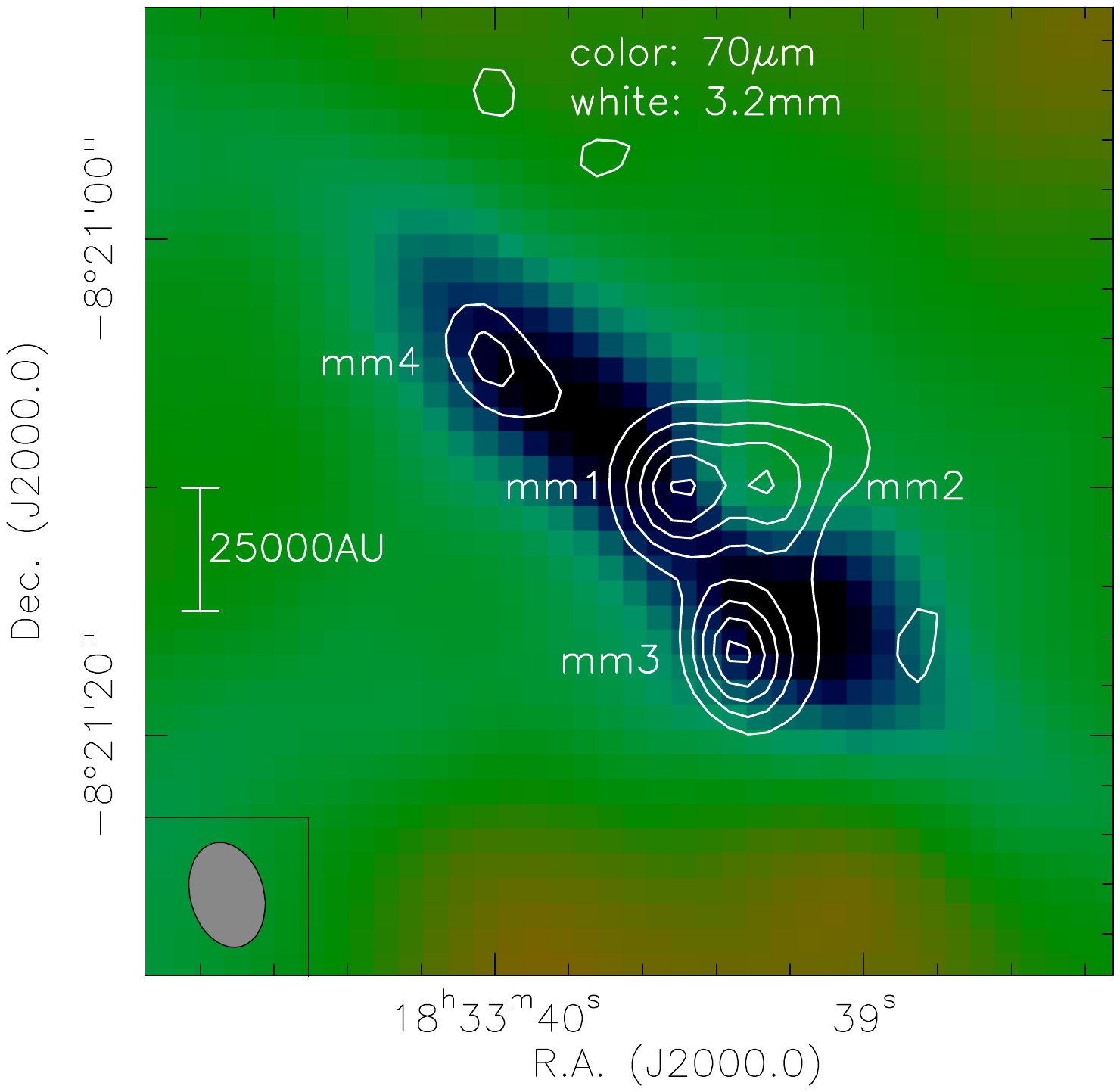}
\caption{Zoom into the central IRDC\,18310-4. The color-scale shows
  the 70\,$\mu$m extinction feature, and the contours present the
  3.2\,mm PdBI continuum emission starting at 3$\sigma$ and continuing
  in $2\sigma$ steps ($1\sigma=0.08$\,mJy\,beam$^{-1}$).  The
  mm-labels, a synthesized beam and a scale-bar are shown as well.
  The phase reference center is close to mm1 at R.A.~(J2000.0)
  18:33:39.468 and Dec.~(J2000.0) -08:21:07.65, and the primary beam
  of the observations with $\sim 54''$ is larger than the shown image
  size.}
\label{zoom}
\end{figure}

\subsection{APEX and the IRAM 30\,m continuum data}

The 1.2\,mm continuum data were first presented in
\citet{beuther2002a} and the APEX 870\,$\mu$m data are part of the
ATLASGAL survey of the Galactic plane \citep{schuller2009}. Beam sizes
of the two datasets are $10.5''$ and $19.2''$, respectively. The
$1\sigma$ rms values are 12 and 63\,mJy\,beam$^{-1}$, respectively.

\section{Results}

\subsection{Continuum emission}
\label{continuum}

Figure \ref{overview} presents an overview of the far-infrared to
submm continuum data obtained toward the whole cloud complex with
Herschel and APEX. While the IRAS source and an additional
far-infrared source in the north are bright at all presented
wavelengths, our target region IRDC\,18310-4 shows considerable
structure changes between 70\,$\mu$m and 870\,$\mu$m. While it is a
strong emission source at long wavelengths, the source is only
depicted as an absorption shadow at 70 and 100\,$\mu$m. While we
cannot exclude that some ongoing low-mass star formation may exist
below our detection threshold, non-detections at far-infrared wavelengths
are among the best indicators for genuine high-mass starless gas
clumps.

\begin{table}[htb]
\caption{Projected nearest neighbor separations}
\begin{tabular}{lrr}
\hline \hline
 & ('') & (AU) \\
\hline
mm1--mm2 & 2.9 & 14200 \\
mm1--mm3 & 7.1 & 34900 \\
mm1--mm4 & 9.5 & 46500 \\
\hline \hline
\end{tabular}
\label{separations}
\end{table}

The longest wavelengths 1.2\,mm continuum data with a spatial
resolution of $10.5''$ first presented in \citet{beuther2002a} are
best suited to re-estimate the gas mass and column densities of
IRDC\,18310-4. Peak and integrated 1.2\,mm fluxes are
132\,mJy\,beam$^{-1}$ and 717\,mJy for that region, respectively.
Assuming optically thin thermal emission from dust, a gas-to-dust mass
ratio of 186 \citep{jenkins2004,draine2007} and a temperature of 18\,K
derived from NH$_3$ observations \citep{sridharan2005} as well as the
Herschel continuum data (see below), we can estimate masses and column
densities for different dust properties.  Using standard ISM dust
properties (\citealt{mathis1977}, MRN1977, ($\kappa_{\rm{1.2mm}}\sim
0.4$\,cm$^2$g$^{-1}$)), the clump mass and peak column density are
$\sim$1600\,M$_{\odot}$ and $2.5\times 10^{23}$\,cm$^{-2}$,
respectively. If we use instead the dust properties discussed in
\citet{ossenkopf1994} for thin ice mantles at densities of
$10^5$\,cm$^{-3}$ (OH94, $\kappa_{\rm{1.2mm}}\sim
0.8$\,cm$^2$g$^{-1}$), the corresponding values for mass and column
density are $\sim$800\,M$_{\odot}$ and $1.3\times 10^{23}$\,cm$^{-2}$,
respectively. These two estimates can be considered as brackets around
the real gas masses and column densities of that region.

\begin{figure}[htb] 
\includegraphics[width=9cm]{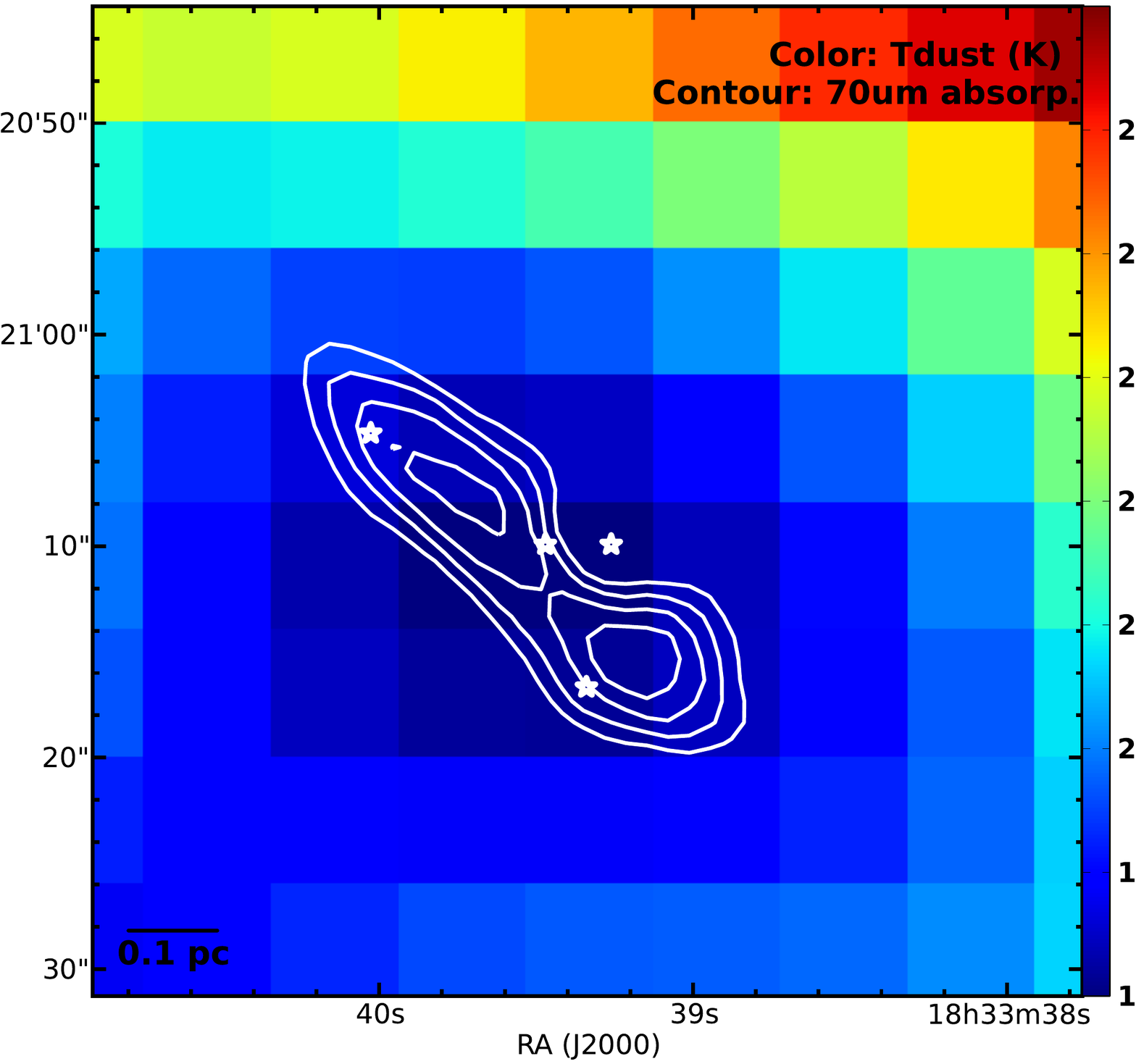}
\caption{The color-scale presents the dust temperature map derived
  from the Herschel far-infrared continuum data. Contours outline the
  structure of the 70\,$\mu$m extinction, and the stars mark the
  positions of the four mm-sources from Figure \ref{zoom}.}
\label{tdust}
\end{figure}

Assuming a standard initial mass function (e.g., \citealt{kroupa2001})
and a relatively high star formation efficiency of $\sim 30\%$ which
should be appropriate for such dense gas clumps (e.g.,
\citealt{alves2007}), one needs approximately 1000\,M$_{\odot}$
initial gas mass to form a cluster with at least one 20\,M$_{\odot}$
star in the end. Given the mass ranges estimated above, IRDC\,18310-4
should be capable of high-mass star formation. Considering column
density thresholds, \citet{tackenberg2012} recently estimated that the
initial gas column density for the Orion Nebula Cluster should have
been approximately $1.8\times 10^{23}$\,cm$^{-2}$, which is roughly
agrees with the theoretical column density threshold from
\citet{krumholz2008b}. Comparing this to the peak column density
derived from the 1.2\,mm data above, it also indicates that this
region may form high-mass stars in the future. \citet{commercon2011}
recently pointed out that not only high column density and heating can
prevent fragmentation but that the magnetic field may be as important.

%

Figure \ref{zoom} presents a zoom toward our central target region
IRDC\,18310-4. With the PdBI 3.2\,mm continuum data, we can identify
four sub-sources (mm1 to mm4, in the following also called cores, in
contrast to the larger-scale environment labeled clump) above a
$5\sigma$ threshold. While we detect a few more $3\sigma$ features,
for the following analysis we only consider the $5\sigma$ detections
mm1 to mm4. The cores are marginally resolved, and the source
separation between mm1 to mm3 is done ``by eye'' at the lowest flux
decrements between the cores. Outer edges are chosen at the $3\sigma$
level.  Peak and integrated fluxes for these four cores are presented
in Table \ref{fluxes}. Using the dust emission again to estimate gas
masses and column densities with the same assumptions outlined above
(MRN1977, $\kappa_{\rm{3.2mm}}\sim 0.06$\,cm$^2$g$^{-1}$), OH94,
$\kappa_{\rm{3.2mm}}\sim 0.17$\,cm$^2$g$^{-1}$), for these four cores
we find masses between 18 and 107\,M$_{\odot}$ and peak column
densities between $1.1\times 10^{23}$ and $6.1\times
10^{23}$\,cm$^{-2}$, respectively (Table \ref{fluxes}).  While we
obviously filter out a significant fraction of the flux with the
interferometer ($\sim 73\%$), this region still hosts relatively
massive cores within a high-mass gas clump reservoir.

While the locations of mm1, mm3 and mm4 are largely associated with
the 70\,$\mu$m extinction feature (although mm1 and mm3 are both
shifted a bit to the edge), it is surprising that mm2 lies at the edge
of the dark extinction feature at a narrowing of the extinction.
Although we cannot identify a clear and isolated 70\,$\mu$m or shorter
wavelength point source at this position, the slight increase of
70\,$\mu$m emission toward the mm2 peak is consistent with a very
young embedded object that just started to heat up the environment. In
this picture, it may be that mm2 is in a little more advanced
evolutionary stage compared to the other mm peak positions.

Another characteristic of the fragmentation of the gas clump is the
projected nearest neighbor separation which is listed in Table
\ref{separations}. Since the beam is elongated in approximately in
north-south direction, we can resolve in east-west direction a
separation between mm1 and mm2 of about $\sim 2.9''$, which
corresponds at the given distance of 4.9\,kpc to a projected linear
separation of of $\sim 14200$\,AU. The projected nearest neighbor
separations to the other cores is even larger, between $\sim 35000$
and $\sim 47000$\,AU (Table \ref{separations}). In comparison to these
values, the approximate Jeans-length where gas clumps should fragment
into sub-structures at approximate temperatures and densities of 18\,K
and $10^5$\,cm$^{-3}$ is $\sim 17000$\,AU.  Hence, we see
fragmentation above and below this approximate Jeans-length which is
reasonable since gas clumps have a density structure and therefore not
one unique fragmentation scale.

We also use the Herschel data to derive the temperature structure of
the region fitting the SEDs pixel by pixel (see also
\citealt{stutz2010} for details of the fitting). We excluded the
500\,$\mu$m data to achieve a higher spatial resolution for the
temperature map of $24.9''$, corresponding to the beam of the Herschel
350\,$\mu$m data.  Figure \ref{tdust} presents the results and we see
a clear drop of the far-infrared dust temperature from the outside to
the center of our region where temperatures around 18\,K are measured.
This is very similar to the gas temperature derived toward that region
by NH$_3$ observations \citep{sridharan2005}, and it is further
evidence that no internal heating source exists so far.

\begin{figure}[htb] 
\includegraphics[width=11cm]{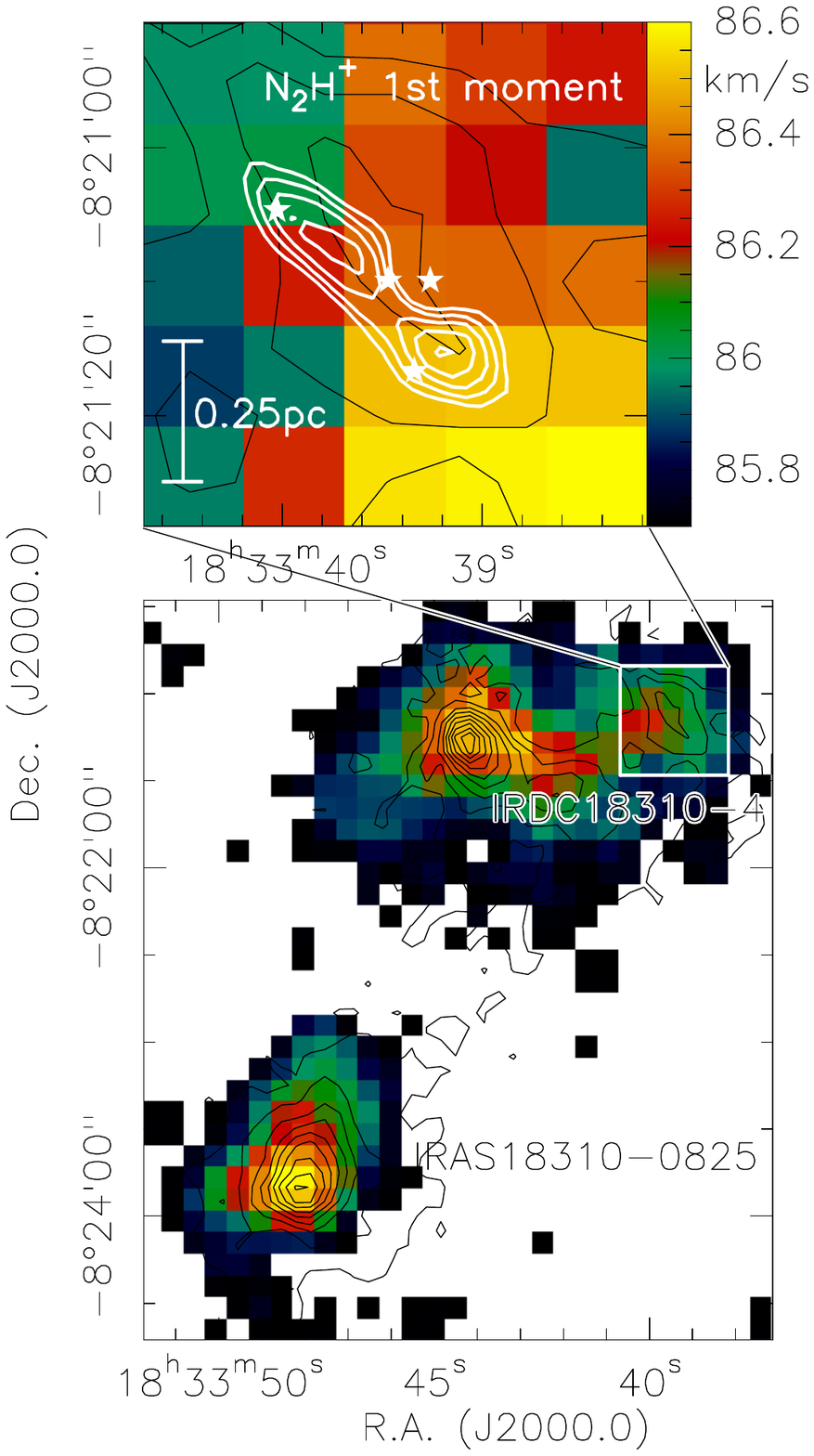}
\caption{The bottom panel shows in color the integrated
  N$_2$H$^+(1-0)$ emission (the whole hyperfine structure between 76
  and 94\,km\,s$^{-1}$) observed with the Nobeyama 45\,m telescope
  toward the entire complex containing IRAS\,18310-0825 in the
  south-east and IRDC\,18310-4 in the north-west. The contours show
  the 1.2\,mm continuum data presented first by \citet{beuther2002a}.
  The contouring is from 10 to 90\% of the peak emission of
  400\,mJy\,beam$^{-1}$. The top panel shows a zoom into IRDC\,18310-4
  where the color scale presents a 1st moment map (intensity-weighted
  peak velocity) of the main hyperfine component complex between 84
  and 90\,km\,s$^{-1}$. The black contours are the same 1.2\,mm map as
  in the bottom panel, and the white contours outline the 70\,$\mu$m
  extinction feature. The stars mark the positions of the four PdBI
  3.2\,mm sources.}
\label{single-dish}
\end{figure}

\subsection{Spectral line emission}
\label{lines}

To study the kinematics of the gas, Figure \ref{single-dish} presents
the N$_2$H$^+(1-0)$ data (critical density at 20\,K is $1.6\times
10^5$\,cm$^{-3}$) of the region observed by Tackenberg et al.~(in
prep.). Although the N$_2$H$^+$ emission toward IRDC\,18310-4 is a bit
weaker than toward the HMPO IRAS\,18310-0825, we nevertheless see a
clear N$_2$H$^+$ peak toward our target region. Zooming into
IRDC\,18310-4, the top panel of Figure \ref{single-dish} shows a 1st
moment map (intensity-weighted peak velocity) extracted from these
data using only the main hyperfine structure peak (see also
Fig.~\ref{spec_nobeyama}). This figure shows on the one hand that the
overall velocity structure of the region is relatively uniform with no
pronounced velocity structure and only weak gradients across the whole
region ($<1$\,km\,s$^{-1}$ over $\sim$1\,pc extent).

Figure \ref{spec_nobeyama} shows a spectrum extracted toward the peak
position. Observing the whole N$_2$H$^+(1-0)$ hyperfine structure
simultaneously allows us to fit the optical depth and by that derive
accurate spectral line parameters. For the single-dish spectrum toward
the peak position, we find a line-width of 2.7\,km\,s$^{-1}$ at a
velocity of 86.1\,km\,s$^{-1}$.  The fitted peak velocity is very
close to that measured previously in NH$_3$ (86.5\,km\,s$^{-1}$,
\citealt{sridharan2005}).  Following \citet{maclaren1988}, we can
estimate a virial mass assuming the FWHM/2 of the Nobeyama beam as the
radius of the core. Employing possible density structures of $1/r$ and
$1/r^2$ as border conditions, the virial mass of this region is
between 305 and 202\,M$_{\odot}$, respectively. Compared to the total
gas masses derived from the single-dish dust continuum data described
in the previous section (section \ref{continuum}), the virial mass is
considerably lower (between factors 2.6 and 7.7, depending on the
assumptions for the dust properties and density structure). This
implies that the whole region is gravitationally unstable on these
scales and likely already undergoing collapse.

\begin{figure}[htb] 
\caption{Compilation of N$_2$H$^+(1-0)$ Nobeyama 45\,m single-dish
  spectrum toward IRDC\,18310 (black histogram with green fit) and
  PdBI spectrum toward mm1 (red histogram). The PdBI data are
  multiplied by 5 for clarity. Fit results to the whole
  N$_2$H$^+(1-0)$ hyperfine structure are given in the panel.}
\label{spec_nobeyama}
\end{figure}

Since our PdBI observations are not only sensitive to the continuum
but also to the N$_2$H$^+(1-0)$ emission, we have kinematic
information at high spatial resolution as well. 
Figure \ref{spec_nobeyama} also shows an N$_2$H$^+(1-0)$ spectrum
extracted toward the peak of mm1 just from the PdBI data (scaled up by
a factor 5), and one clearly sees kinematic differences between the
large and the small scale. To account for the differences and to
complement the PdBI data with the missing flux information, we merged
the PdBI and Nobeyama data into one single data-cube (see Section
\ref{mergedata}).

Figure \ref{pdbi_nobeyama} presents the spectra extracted from the
merged data-cube toward the mm peak positions. All these spectra show
that we have at least two separated velocity components in the region,
one approximately peaking at $\sim$86.5\,km\,s$^{-1}$ and the other at
$\sim$89.3\,km\,s$^{-1}$ (these values are averages from the full
hyperfine structure fits to mm1 to mm3, see Fig.~\ref{pdbi_nobeyama}).

\begin{figure}[htb] 
\caption{Spectra and N$_2$H$^+(1-0)$ hyperfine spectral line fits
  extracted from the PdBI+Nobeyama N$_2$H$^+(1-0)$ data toward the
  mm-peak positions marked in Figure \ref{zoom}. Fit results and
  source labels are presented in each panel.}
\label{pdbi_nobeyama}
\end{figure}

\begin{figure*}[htb] 
\caption{PdbI only (left panel) and merged PdBI+Nobeyama 45\,m
  N$_2$H$^+(1-0)$ data (right panel). The color-scale shows the
  70\,$\mu$m extinction feature, and the contours present the
  integrated N$_2$H$^+(1-0)$ emission of three different velocity
  components as indicated in the figure. To avoid hyperfine-splitting
  overlap, these integrated maps were created from the isolated
  hyperfine component $-8.0$\,km\,s$^{-1}$ apart from the central
  component, and then the velocity shifted to the center to be
  comparable with other line measurements. The contouring is done from
  15 to 95\% of the peak emission of each dataset, respectively.  The
  mm-labels, a synthesized beam and a scale-bar are shown as well.}
\label{merge}
\end{figure*}

The comparison of the PdBI only spectrum and the Nobeyama only
spectrum in Figure \ref{spec_nobeyama} shows that the lower-velocity
PdBI component corresponds relatively well to the single-dish spectrum
whereas the higher-velocity PdBI component does not have a proper
single-dish spectrum counterpart. This strongly indicates that the
$\sim$86.5\,km\,s$^{-1}$ component is the main component also present
in the ambient gas, whereas the higher velocity component at
$\sim$89.3\,km\,s$^{-1}$ appears to be mainly associated with a dense
sub-part moving with respect to the ambient gas.

To visualize spatial differences between these two spectral
components, we imaged the N$_2$H$^+$ data for both velocity structures
separately. Figure \ref{merge} shows the integrated emission derived
for the velocity regimes [85.5,88.5] and [88.5,89.7]\,km\,s$^{-1}$,
respectively. The left panel shows the PdBI- only data whereas the
right panel presents the combined PdBI+Nobeyama 45\,m data.  To avoid
any artifacts of overlapping hyperfine structure lines, we produced
these images from only the isolated hyperfine structure component
offset by $-8.0$\,km\,s$^{-1}$ ($F=0-1$). While the lower-velocity
component agrees well with the mid-infrared absorption structure
encompassing mm1, mm3 and mm4, the higher-velocity component is
peaking mainly in the direction of mm2. It is at first sight
surprising to note that the higher-velocity component in the PdBI-only
data shows an extension toward mm4 whereas this is lost in the merged
dataset. Most likely this is because that extension is on small scales
and comparably weak that it gets lost by the merging process.
Additionally, it is possible that the single-dish data add more noise
in the combined dataset which make this weaker extension harder to
detect. In addition to this, the merged dataset shows one additional
velocity component at even higher velocities between 89.7 and
90.6\,km\,s$^{-1}$ that is more diffusely distributed north of the
main IRDC\,18310-4 structures. It is unclear whether this is just
another unrelated gas component or whether that gas is structurely
related to the red-shifted component between 88.5 and
89.7\,km\,s$^{-1}$. Nevertheless, both main velocity components
strongly overlap and hence both contribute to the N$_2$H$^+$ emission
from mm1 and mm2 (see spectra in Fig.~\ref{pdbi_nobeyama}).
Furthermore, the PdBI-only data indicate that mm4 is located at an
interface region between both components. In contrast to this, mm3
appears a bit different because it is located at the edge of the
integrated emission of both components.

Table \ref{fluxes} lists the line widths of the two components as well
as the estimated virial masses following \citet{maclaren1988} for
different density profiles ($\rho\propto \frac{1}{r}$ and $\rho\propto
\frac{1}{r^2}$). The assumed radius of the sources is 1/2 of the FWHM
which corresponds at the given distance to $\sim 9000$\,AU. It is
obvious that the narrow higher-velocity component barely contributes
to the stability of the cores, and that the broader low-velocity
component is far more important to stabilize the structures.
Considering the spread in virial masses for different density profiles
as well as for the gas masses based on the dust continuum emission and
the assumed different dust properties (Table \ref{fluxes}), the cores
appear closer to virial equilibrium. This is different to the overall
virial to gas mass ratio based on the single-dish data for the whole
region that is significantly smaller than 1 (see above).  However, one
has to keep in mind that the virial masses are calculated from the
merged PdBI+Nobeyama N$_2$H$^+$data whereas the core masses are
calculated from the PdBI data only. If one compares the main component
of the N$_2$H$^+$ spectrum toward mm1 without and with the Nobeyama
short spacing information (Figs.  \ref{spec_nobeyama} \&
\ref{pdbi_nobeyama}), one finds that the combined data reveal a FWHM
about a factor 1.7 higher than the interferometer data only. Since the
line-widths is squared in the estimate of the virial mass, the derived
virial masses of the combined data are usually a factor of a few
larger than with interferometer data only. This implies that the
comparison of masses from the combined PdBI+Nobeyama line data with
the PdBI-only continuum data is not reliable. With the missing flux
problem, the core masses from the PdBI-only data are underestimated,
implying that the ratio of virial to gas mass on the core scales
should be smaller than 1 as well. Hence, the cores are also prone to
collapse.

Figure \ref{moments} presents the integrated N$_2$H$^+$ emission as
well as the first and second moment maps (intensity-weighted peak
velocities and line widths). In particular the 1st moment map shows
that we have no strong velocity gradients across the filament but
that the velocity peaks are dominated by the two components discussed
above.  The 2nd moment map appears as if it had a line width increase
toward mm2, however that is only because the 2nd moment algorithm
assumes a single component, and therefore the two real components
appear as one broader feature in that direction.

\begin{figure*}[htb] 
\caption{N$_2$H$^+(1-0)$ moment maps of the PdBI-only data toward
  IRDC\,18310-4. The left, middle and right panel show in color the
  0th moment (integrated intensity), 1st moment (intensity-weighted
  peak velocity) and 2nd moment (intensity-weighted linewidth),
  respectively. The moments were again extracted for the isolated
  hyperfine component $-8.0$\,km\,s$^{-1}$ from the line center.  The
  1st moment was calculated from this isolated hyperfine component and
  then shifted toward by $+8.0$\,km\,s$^{-1}$ to the $v_{lsr}$ of the
  source again. The white contours outline the 70\,$\mu$m extinction
  feature and the stars mark the 3.2\,mm peak positions from Figure
  \ref{zoom}. The synthesized beam and a scale-bar are shown again.}
\label{moments}
\end{figure*}

\section{Discussion}

\subsection{Linear scales and densities from low- to high-mass
  starless gas clumps}
\label{jeans}

\begin{figure*}[htb] 
\caption{Comparison between starless gas-clumps from low- via
  intermediate- to high-mass star formation. The 875\,$\mu$m data for
  the prototypical low-mass core B68 are taken from
  \citet{nielbock2012}, and the N$_2$H$^+$(1--0) data for the
  intermediate-mass star-forming region IRDC\,19175 are from
  \citet{beuther2009b}. Each panel presents a linear scale-bar of
  50000\,AU and the synthesized beam, average densities are mentioned
  as well.}
\label{compare}
\end{figure*}

How do linear scales and average densities compare between
star-forming regions from low- to high-mass? To compare the different
mass regimes with the high-mass starless region presented in this
paper, we selected the low-mass starless core B68 (e.g.,
\citealt{alves2001,nielbock2012}) which resembles typical other
low-mass globules (e.g., \citealt{launhardt2010}), and the
intermediate-mass starless star-forming region IRDC\,19175
\citep{beuther2009b}. Figure \ref{compare} presents data for all three
regions. It is interesting to note that all three regions have
approximately the same linear spatial extent on the order of
50000\,AU. However, while the linear size between the three regions
does not vary considerably, the total gas reservoirs for the three
complexes are very different, ranging from $\sim$3.1\,M$_{\odot}$ for
the low-mass core B68 \citep{nielbock2012}, $\sim$87\,M$_{\odot}$ for
the intermediate-mass region IRDC\,19175 \citep{beuther2009b} to
values between $\sim$800 and $\sim$1600\,M$_{\odot}$ (depending on the
dust model, see section \ref{continuum}) for the high-mass region
IRDC\,18310-4 (see Table \ref{parameters} for a list of the source
parameters).  Assuming that the three regions resemble typical
starless gas clumps in the different mass regimes, these data imply
that not the linear sizes are important but the mass reservoir
squeezed within such regions, i.e., the densities.


While the peak density of the B68 core is $\sim 3.4\times
10^5$\,cm$^{-3}$ \citep{nielbock2012}, the average density of the
region is considerably lower. We estimate an average density for B68
from the total mass of 3.1\,M$_{\odot}$ (recent analysis of Herschel
data, \citealt{nielbock2012}) divided by a sphere with approximate
radius of $\sim 150''$ at a given distance of $\sim 150$\,pc. Using
these numbers the average gas density of the low-mass starless core
B68 is approximately $1.2\times 10^4$\,cm$^{-3}$ over a region
corresponding to a sphere with $\sim$45000\,AU diameter.

For the intermediate-mass star-forming region IRDC\,19175, average
core densities are estimated by \citet{beuther2009b} to $\sim 6\times
10^5$\,cm$^{-3}$ over regions corresponding to spheres with
$\sim$3500\,AU diameter. Estimating an average density from the total
mass of 87\,M$_{\odot}$ over a volume corresponding to an effective
radius of the 1.2\,mm continuum map in \citet{beuther2009b} results
only in slightly lower values around $\sim 2.6\times 10^5$\,cm$^{-3}$
over a region corresponding to a sphere with $\sim$48000\,AU diameter.

To compare these values with our high-mass star-forming region
IRDC\,18310-4, we estimate an average density for the core mm1 via
averaging the peak column densities listed in Table \ref{fluxes} for
the two different dust models, and assuming that the 3rd dimension has
the same size as our synthesized beam. This results in average
densities for mm1 of $\sim 1.5\times 10^6$\,cm$^{-3}$ over a region
corresponding to a sphere with $\sim$18000\,AU diameter. To get the
density over a slightly larger regions, we use the average column
density derived in section \ref{continuum} for the two different dust
models from the single-dish 1.2\,mm continuum data and again assume
the 3rd dimension to have the same spatial extent as the single-dish
beam of $10.5''$. This results in average densities of $\sim 2.5\times
10^5$\,cm$^{-3}$ over a region corresponding to a sphere with
$\sim$51000\,AU diameter.

\begin{table}[htb]
\caption{Star-forming region parameters}
\begin{tabular}{lrrr}
\hline \hline
 & B68 & IRDC19175 & IRDC18310-4 \\
\hline
mass [M$_{\odot}$] & 3.1 & 87 & 800--1600 \\
\hline
av.~dens.~[cm$^{-3}$] & 1.2$\times 10^4$ & 2.6$\times 10^5$ & 2.5$\times 10^5$\\
size [AU] & 45000 & 48000 & 51000\\
Jeans length$^2$ [AU] & 44000 & 10000 & 10000\\
Jeans mass$^2$ [M$_{\odot}$] & 1.7 & 0.36 & 0.37 \\
\hline
core dens.~[cm$^{-3}$] & --$^1$ & 6.0$\times 10^5$ & 1.5$\times 10^6$\\
Jeans length$^2$ [AU] & --$^1$ & 6000 & 4000\\
Jeans mass$^2$ [M$_{\odot}$] & --$^1$ & 0.24 & 0.15 \\
\hline \hline
\end{tabular}
\footnotesize{~\\ $^1$ B68 as a whole is already the core.\\
$^2$ At 15\,K}
\label{parameters}
\end{table}

These large density differences obviously also have a strong impact on
the fragmentation properties of the starless gas clumps. While the
low-density low-mass core B68 does not fragment at all, the
higher-density intermediate- to high-mass starless clumps do fragment
significantly. To better understand these properties, we estimate the
Jeans length for the average B68 density of $1.2\times
10^4$\,cm$^{-3}$, the approximate average large-scale densities of
IRDC\,19175 and IRDC\,18310-4 of $\sim 2.5\times 10^5$\,cm$^{-3}$, and
the higher core densities of the two regions of $\sim 6\times
10^5$\,cm$^{-3}$ and $\sim 1.5\times 10^6$\,cm$^{-3}$, respectively.
Assuming average temperatures of 15\,K for all these starless regions,
the corresponding Jeans-length are $\sim$44000\,AU (B68),
$\sim$10000\,AU (average large-scale densities of IRDC\,19175 \&
IRDC\,18310-4), $\sim$6000\,AU (average core density of IRDC\,19175)
and $\sim$4000\,AU (average core density of IRDC\,18310-4),
respectively. This relatively simple isothermal Jeans-analysis already
explains the observations well: The large Jeans-length for B68 is
reflected in the fact that this low-mass starless core does not
fragment on the given scales (Fig.~\ref{compare}). Furthermore, as
already discussed in \citet{beuther2009b}, the different Jeans-lengths
for B68 (located close to the Pipe nebula) and IRDC\,19175 correspond
well to the observed core separations in the low-mass Pipe nebula and
the intermediate-mass IRDC\,19175 region, respectively. For the
high-mass region IRDC\,18310-4, we are still spatial resolution
limited and only observe the fragmentation on the scales of our
observed resolution elements (see also Table \ref{separations}). In
future higher-spatial-resolution observations, we expect to resolve
this region into even more and closer spaced cores.

Calculating the Jeans-mass instead of the Jeans-length for the average
densities of B68, IRDC\,19175 and IRDC\,18310-4, we find 1.7, 0.36 and
0.37\,M$_{\odot}$, respectively. The Jeans-masses for the higher core
densities are correspondingly smaller (Table \ref{parameters}). For
the low-mass region B68 that value is on the order of the whole mass
(3.1\,M$_{\odot}$), and also for the intermediate-mass region
IRDC\,19175, the estimate masses of the individual cores are mostly
below 1\,M$_{\odot}$. Hence Jeans-masses and core masses approximately
agree in these two cases. In contrast to that, the Jeans-mass in
IRDC\,18310-4 is significantly lower than the mass estimates we
derived from the dust continuum data (Table \ref{fluxes}). Therefore,
similar as discussed in the previous paragraph, for the high-mass
region IRDC\,18310-4, we are still spatial resolution limited, and it
is highly likely that the here presented cores mm1 to mm4 will
fragment even further when observed at higher spatial resolution.

Assuming that these example regions resemble starless gas clumps over
the whole mass regime, one may conclude that gravitational instability
in the classical Jeans-analysis explains the fragmentation properties
of starless gas clumps from low- to high-mass reasonably well. The
major difference between the regions appears to be their densities and
mass reservoirs squeezed into the respective spatial scales. Similar
Jeans fragmentation scales have also been observed toward more evolved
regions, e.g., the young cluster NGC2264 \citep{teixeira2006}. In
contrast to the larger-scale cloud fragmentation where turbulence
plays a major role (e.g., \citealt{padoan2002,maclow2004}), these data
indicate that the fragmentation on smaller clump and core scales
appears to be dominated by gravity.

\subsection{Kinematic signatures of a dynamic cloud
  collapse}
\label{discussion}

Independent of the cloud formation models -- do clouds evolve slower
via ambipolar diffusion processes (e.g.,
\citealt{mouschovias2006,mckee2007}) or faster via more dynamic gas
flows (e.g., \citealt{vazquez2006,heitsch2008,banerjee2009}) -- the
gas has to show dynamic signatures of contraction and/or infall over
large spatial scales. During molecular cloud formation the diffuse HI
gas converts to the denser molecular phase. Within molecular clouds,
filamentary sub-structures are likely important, and recent data
indicate that hierarchical, often also filamentary structures feed the
central and densest filaments or gas clumps (e.g.,
\citealt{myers2009,schneider2010,schmalzl2010,hennemann2012}). These
central cluster-forming clumps have then to collapse and form the
multiple and clustered stellar associations in the end. Statistical
signatures of infall from single-dish average line profiles exist.
However, they are often ambiguous, and one finds signatures for infall
as well outflowing motions (e.g., \citealt{fuller2005}). More coherent
infall motions on larger scales have only rarely been reported (e.g.,
\citealt{motte2005,schneider2010}). Going from the cluster-forming
scales down to cores forming individual or multiple, bound
objects, dynamic studies in the past have mostly concentrated on more
evolved evolutionary stages like high-mass protostellar objects
(HMPOs) or ultracompact H{\sc ii} regions (e.g.,
\citealt{keto2002a,sollins2005,sandell2009,qiu2011,goddi2011,beuther2012c}).

While the above infall studies dealt with the average infall motions
of the entire star-forming gas clumps, only recently spatially
resolved dynamical studies of sub-structures and cores within the
natal gas became possible (e.g.,
\citealt{beuther2009b,zhang2011,wang2011,csengeri2011a,csengeri2011b}).
In contrast to the above discussed single-dish studies where often
infall signatures are derived from line profiles of optically thick
lines, \citet{ragan2012a} also find signatures of collapse and
fragmentation via high-spatial-resolution NH$_3$ observations of
IRDCs. Furthermore, the studies by \citet{csengeri2011a,csengeri2011b} revealed
multiple velocity components in optically thin gas tracers like
H$^{13}$CO$^+$ or N$_2$H$^+$ that were interpreted as signatures of
shear motions possibly caused by converging gas flows. 

The spectra we found (Figs.~\ref{spec_nobeyama} and
\ref{pdbi_nobeyama}) also show multiple velocity components in the
dense gas tracer N$_2$H$^+$. Fig.~\ref{merge} even identifies that the
different velocity components are dominated by different mm cores
within the large-scale gas clump.  In addition to the work by
\citet{csengeri2011a,csengeri2011b}, multiple velocity components in
optically thin dense gas tracers were also found by
\citet{beuther2009b} as well as in recent work by Ragan et al.~(in
prep.) with the PdBI toward the G11.11 filament, and by Bihr et
al.~(subm.) with VLA NH$_3$ observations toward IRDCs. Combining
our new data with the results recently found by the other groups,
based on high-spatial resolution interferometer observations there
emerges a regular signature of multiple velocity components in
optically thin dense gas tracers. Typical separations line widths of
these multiple velocity components are around $\sim$2\,km\,s$^{-1}$
and $\sim$1\,km\,s$^{-1}$, respectively.

\begin{figure}[htp]
\begin{center}
\caption{The N$_2$H$^+$ (1-0) isolated hyperfine line profiles viewed
  through various lines of sight through a massive star forming
  region. The central image with a side-length of 0.8\,pc shows the
  column density of the region and the surrounding line profiles are
  calculated for a 0.09\,pc beam centered directly on the embedded
  core and viewed along the direction each panel touches the central
  image. }
\label{sightlines}
\end{center}
\end{figure}

Confronting these observational results with theory, Rowan Smith has
recently started to simulate the spectral line emission from
dynamically collapsing clumps of a large-scale collapsing gas
cloud modeled initially by \citet{smith2009a} and \citet{bonnell2011}.
First results about cores associated with low-mass star formation were
presented in \citet{smith2012}, and the simulations for high-mass
collapsing gas clumps are currently conducted (Smith et al.~subm.).
Figure \ref{sightlines} shows an example of non-Gaussian optically
thin line profiles produced from a simulation of a massive
star-forming gas clump with 370\,M$_{\odot}$ (\citealt{smith2009a},
Smith et al.~subm.). The N$_2$H$^+$ (1-0) line profiles were
calculated using an assumed abundance ratio of $10^{-10}$ relative to
molecular hydrogen \citep{aikawa2005}. The radiative transfer code
RADMC-3D \citep{dullemond2012} was used to calculate the line emission
given the temperatures and densities from the original simulation.
(For a full description of our method see \citealt{smith2012}). Here,
only the isolated hyperfine component $F=0-1$ (see
Fig.~\ref{pdbi_nobeyama}) offset by $\sim$8.0\,km\,s$^{-1}$ from the
central component is shown.  The line profiles are synthesized over a
beam of FWHM 0.09\,pc and calculated for a variety of viewing angles.
No artificial noise is added to the data.

Clearly, these simulations find similar signatures of multiple
N$_2$H$^+$ components within the dynamically collapsing gas clump.
Typical peak separations and line widths are similar to what is found
in the observational data. The multiple components seen in Figure
\ref{sightlines} arise in the simulations due to the clumpy nature of
the high-mass star-forming region. Across the region there is a large
scale supersonic collapse motion running from around +2\,km\,s$^{-1}$
on one side to $-2$\,km\,s$^{-1}$ on the other. Dense clumps of gas at
different positions along this velocity gradient all contribute to the
total emission and produce peaks in the line profile at their
individual velocities. A full analysis of line profiles from massive
star formations will be presented in Smith et. al.(subm.).

This interpretation is also consistent with our finding that the ratio
of viral mass to total gas mass is significantly smaller than 1 for
the entire region, implying that the large-scale gas clump is prone to
collapse, similar to the findings of \citet{ragan2012a} for their
sample of IRDCs. Since the virial mass to core mass ratio is smaller
than 1 on the core scales as well, the region is consistent with a
globally collapsing gas clumps on almost all scales.

We can also estimate the dynamical crossing time as well as the
free-fall time of the region. Using the typical velocity difference
between the two gas components of $\sim 2.8$\,km\,s$^{-1}$ as the
velocity dispersion of the clumps, and the approximate extent of the
70\,$\mu$m absorption feature of $\sim 30''$ (corresponding to
147000\,AU at the given distance), the dynamical crossing time for
IRDC\,18310-4 is $\sim 2.5\times 10^5$\,yrs. In comparison to that,
estimating the free-fall time for a 1000\,M$_{\odot}$ gas clump of
with a radius of half the previously used extend, we get an
approximate free-fall time-scale of $\sim 1.2\times 10^5$\,yrs.
Although there is a difference of about a factor 2 between the two
estimates, these are just approximate values and they agree
order-of-magnitude wise. This again implies that a dynamical collapse
of the whole gas clump is a feasible model for this region.

Combining our observational results with the modeling of line profiles
for dynamically collapsing gas clumps, we identify signatures that
even this very young, still starless high-mass gas clump is actively
collapsing and thus at the verge of star formation. The fact that
similar signatures were recently found toward other regions as well,
indicates that multiple velocity components of optically thin gas
tracers are an excellent tool to identify and study the earliest
stages of dynamical infall in high-mass star formation.

\section{Conclusions}
\label{conclusion}

Since the evolutionary time-scale for high-mass starless gas clumps is
extremely short (e.g.,
\citealt{motte2007,russeil2010,tackenberg2012}), so far they had
barely been studied in great depth at high spatial resolution (some
notable recent papers in that field are
\citealt{csengeri2011a,ragan2012a}). To better understand the
fragmentation and dynamical properties of the most massive gas clumps
during their earliest evolutionary stages, here we present an
interferometric PdBI kinematic and continuum study of a prototypical
high-mass starless gas clump that remains far-infrared dark up to
100\,$\mu$m wavelengths.

The gas reservoir between $\sim$800 and $\sim$1600\,M$_{\odot}$
(depending on the assumed dust properties) splits up at a spatial
resolution of $4.3''\times 3.0''$ (linear scales $21000\times
15000$\,AU) into 4 cores. A comparison of this high-mass starless
region with typical low-mass (B68) and intermediate-mass (IRDC\,19175)
star-forming forming regions reveals that the overall spatial size of
the low-mass to high-mass star-forming regions does not vary
significantly. While the linear extent of the regions is similar, the
mass reservoir squeezed into this spatial size on the order of
0.25\,pc, and hence the densities do vary by about 2 orders of
magnitude. A classical gravitational instability Jeans analysis is
capable to explain the fragmentation properties of the gas clumps over
the whole mass range presented.

In addition to the fragmentation properties, we find multiple spectral
velocity components toward the resolved gas cores.  Recent radiative
transfer hydrodynamic simulations of dynamical collapsing high-mass
gas clumps are consistent with these multiple velocity components.
This also agrees with a ratio between estimated viral and total mass
of the whole region $<$1.

In summary, although this massive gas clump does not have any embedded
protostellar source down to our Herschel far-infrared detection
limits, the fragmentation and dynamical properties of the gas and dust
are consistent with early collapse motion and clustered star
formation.

\begin{acknowledgements} 
  We like to thank Joao Alves for interesting discussions about the
  comparison between high- and low-mass star-forming regions.
  Furthermore, thanks a lot to a careful referee who helped improving
  the paper.
\end{acknowledgements}


\end{document}